\documentclass[prd,preprintnumbers,twocolumn,eqsecnum,floatfix,letterpaper,superscriptaddress,nofootinbib]{revtex4}
\usepackage{color}
\usepackage{calc}
\usepackage{amsmath,amssymb,graphicx}
\usepackage{amssymb,amsmath}
\usepackage{bm}
\usepackage{times}
\usepackage{microtype}
\usepackage{booktabs}
\usepackage{subfigure}
\usepackage[varg]{txfonts}
\usepackage[colorlinks, pdfborder={0 0 0}]{hyperref}
\definecolor{LinkColor}{rgb}{0.75, 0, 0}
\definecolor{CiteColor}{rgb}{0, 0.5, 0.5}
\definecolor{UrlColor}{rgb}{0, 0, 0.75}
\hypersetup{linkcolor=LinkColor}
\hypersetup{citecolor=CiteColor}
\hypersetup{urlcolor=UrlColor}
\allowdisplaybreaks
\begin{document}

\newcommand{\be}{\begin{equation}}
\newcommand{\ee}{\end{equation}}
\newcommand{\ber}{\begin{eqnarray}}
\newcommand{\eer}{\end{eqnarray}}
\def\bea{\begin{eqnarray}}
\def\eea{\end{eqnarray}}
\newcommand{\ie}{i.e.}
\newcommand{\dt}{{\rm d}t}
\newcommand{\df}{{\rm d}f}
\newcommand{\dtheta}{{\rm d}\theta}
\newcommand{\dphi}{{\rm d}\phi}
\newcommand{\rhat}{\hat{r}}
\newcommand{\iotahat}{\hat{\iota}}
\newcommand{\phihat}{\hat{\phi}}
\newcommand{\hc}{{\sf h}}
\newcommand{\etal}{\textit{et al.}}
\newcommand{\balpha}{{\bm \alpha}}
\newcommand{\bbeta}{{\bm \psi}}
\newcommand{\rmi}{{\rm i}}
\newcommand{\x}{$X$}
\newcommand{\h}{$H$}

\newcommand{\LIGO}{LIGO Laboratory, California Institute of Technology, 
Pasadena, CA 91125, USA}
\newcommand{\CIT}{Theoretical Astrophysics, California Institute of
Technology, Pasadena, CA 91125, USA}
\newcommand{\ICTS}{International Centre for Theoretical Sciences, 
Tata Institute of Fundamental Research, Bangalore 560012, India.}
\newcommand{\MIT}{LIGO Laboratory, Massachusetts Institute of Technology, Cambridge, MA 02139 USA}
\newcommand{\Carleton}{Physics and Astronomy, Carleton College, Northfield, MN 55057, USA}

\title{Instrumental vetoes for transient gravitational-wave triggers using noise-coupling models:\\ The bilinear-coupling veto}
\author{Parameswaran Ajith}
\affiliation{\ICTS}
\affiliation{\LIGO}
\affiliation{\CIT}
\author{Tomoki~Isogai}
\affiliation{\MIT}
\affiliation{\Carleton}
\author{Nelson~Christensen}
\affiliation{\Carleton}
\author{Rana~X.~Adhikari}
\affiliation{\LIGO}
\author{Aaron~B. Pearlman}
\affiliation{\LIGO}
\affiliation{Department of Applied Physics, California Institute of Technology, Pasadena, CA 91125, USA}
\author{Alex~Wein}
\affiliation{\LIGO}
\author{Alan~J.~Weinstein}
\affiliation{\LIGO}
\author{Ben Yuan}
\affiliation{\LIGO}

\begin{abstract}
LIGO and Virgo recently completed searches for gravitational waves at their initial target sensitivities, and soon Advanced LIGO and Advanced Virgo will commence observations with even better capabilities. In the search for short duration signals, such as coalescing compact binary inspirals or ``burst'' events, noise transients can be problematic. Interferometric gravitational-wave detectors are highly complex instruments, and, based on the experience from the past, the data often contain a large number of noise transients that are not easily distinguishable from possible gravitational-wave signals. In order to perform a sensitive search for short-duration gravitational-wave signals it is important to identify these noise artifacts, and to ``veto'' them. Here we describe such a veto, the bilinear-coupling veto, that makes use of an empirical model of the coupling of instrumental noise to the output strain channel of the interferometric gravitational-wave detector. In this method, we check whether the data from the output strain channel at the time of an apparent signal is consistent with the data from a bilinear combination of auxiliary channels. We discuss the results of the application of this veto on recent LIGO data, and its possible utility when used with data from Advanced LIGO and Advanced Virgo.
\end{abstract}
\preprint{LIGO-P1400023-v3}

\maketitle

\section{Introduction}
\label{sec:intro}
The LIGO and Virgo laser interferometric gravitational-wave (GW) detectors recently completed their observations in their initial design configurations. While GWs were not observed, important upper limits have been established in searching for signals from coalescing compact (neutron star and black hole) binaries~\cite{S6inspiral,S6BHinspiral}, burst events~\cite{S6burst} (core collapse supernova~\cite{Ott}, cosmic strings~\cite{S6string}, etc.), rapidly spinning neutron stars~\cite{S5CW}, and a stochastic GW background~\cite{S5stoch}. Searches were also made for GW signals in association with gamma ray bursts~\cite{S6gamma} and high energy neutrinos~\cite{S5neutrinos}. By 2015 Advanced LIGO~\cite{aLIGO} will begin operating with a significant improvement in sensitivity, followed soon thereafter by Advanced Virgo~\cite{aVirgo1,aVirgo2} coming on-line in 2016-2017~\cite{Commissioning}. A world-wide network of advanced interferometric GW detectors will be operating in the near future; a Japanese detector, KAGRA~\cite{KAGRA}, is currently under construction, and a third LIGO detector may also be constructed in India.

Interferometric GW detectors are highly complex instruments; the data to date have often contained a large number of noise transients or noise frequency lines that were not easily distinguishable from possible GW signals. Noise artifacts can be created from imperfections or events within the detector itself, or caused by disturbances in the physical environment around where the detectors are located, which can couple to the output strain channel (the ``GW channel'') through various coupling mechanisms. In order to perform a sensitive search for GW signals, it is important to identify these noise artifacts, and to ``veto'' them. For the initial LIGO and Virgo detectors, numerous techniques were developed in order to identify and remove data from time periods when problems with the detector or its physical environment could be detected~\cite{S6DQ,VSR23DQ}. Similarly, specific noise frequency lines were also identified and removed from searches for GW signals from rapidly spinning neutron stars and the stochastic GW background~\cite{noise-lines}.

Short duration noise transients, or \emph{glitches}, are especially problematic for compact coalescing binary and burst GW signal searches. During the recent LIGO (S6) and Virgo (VSR2, VSR3) scientific runs a number of vetoes were defined in order to identify and remove glitches from the interferometers' output strain GW channel, $H$. During these recent scientific runs data from numerous interferometer auxiliary channels and physical environment monitoring (PEM) devices were recorded, and searched for glitches. The glitch search tool used was a wavelet-based program called \emph{KleineWelle} (KW)~\cite{KleineWelle}. The various vetoes were developed by looking for statistical association between glitches in the interferometer auxiliary channels and the PEM devices and events in the interferometer's output strain channel. For example, the hierarchical (``hveto'') pipeline~\cite{hveto} and the ``used percentage veto''~\cite{UPV} were effective in identifying noise events in the GW channel due to glitches that appeared in multiple channels in LIGO and Virgo data, while the ``SeisVeto''~\cite{MacLeod:2011up} was effective in eliminating glitches that originated due to fluctuations in the seismic noise. Another veto compared KW triggers from the two quadrature phases of Virgo's output strain channel, and when associations could be made between events in the in-phase and quadrature channels, then the in-phase events were vetoed~\cite{PQveto}~\footnote{Another veto method making use of a similar idea, implemented for the GEO\,600 detector, is described in~\cite{Hewitson:2005wr}.}. 

The ``{traditional}'' veto methods mentioned above all search for a time coincidence between a glitch in an interferometer's output strain channel, and an event in an interferometer auxiliary or PEM channel. The \emph{bilinear-coupling veto}, which we are describing in this paper, was developed with the goal to see if the data from an interferometer's output GW strain channel at the time of an apparent signal is consistent with the data from the interferometric detector's auxiliary channels. The consistency check is based on the observation of the coupling of different noise sources to the interferometer output strain channel. This veto was applied on LIGO S6 data~\cite{S6inspiral,S6BHinspiral}, and can be applied on data from Advanced LIGO and Advanced Virgo. In this paper we will fully describe the bilinear-coupling veto, and summarize its results when used on LIGO S6 data. We will also discuss its potential capabilities when used with data from Advanced LIGO and Advanced Virgo.

The organization of the paper is as follows. In Section~\ref{sec:vetomethod} we describe the veto method. The results of the bilinear-coupling veto when applied to LIGO S6 data are given in Section~\ref{sec:veto_analysis_s6}. In Section~\ref{sec:future_work} we discuss how the bilinear-coupling veto can be used as a potential diagnostic tool with the advanced detectors. Concluding observations are given in Section~\ref{sec:conclusions}.

\section{Instrumental veto methods using noise coupling models}
\label{sec:vetomethod}

\begin{figure}[tbh]
\centering
\includegraphics[width=3.4in]{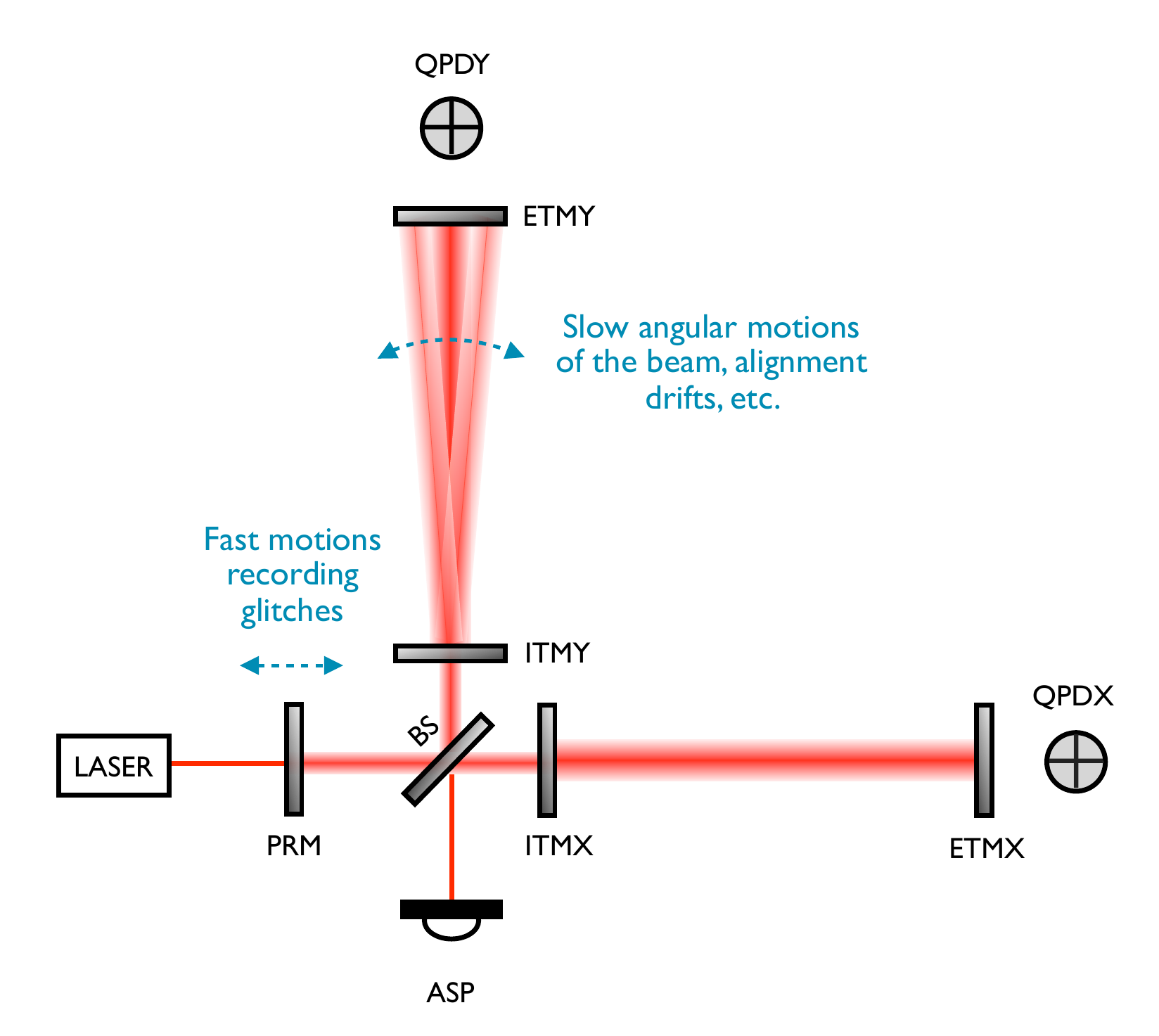}
\caption{Schematic diagram of a power recycled Michelson interferometer.
Examples of motions recorded in ``primary'' and ``secondary'' channels are shown. 
The angular motion shown in the figure is greatly exaggerated. The beam need only move 
a few millimeters away from its optimal position on the mirror to create a non-optimal 
detector ``state''. The label PRM stands for power recycling mirror, BS for beam splitter
and ASP for antisymmetric (``dark'') port; see Section~\ref{sec:veto_analysis_s6} for a 
description of other interferometer components.}
\label{fig:LIGODiagram}
\end{figure}

\begin{figure*}[tbh]
\centering
\includegraphics[width=6.2in]{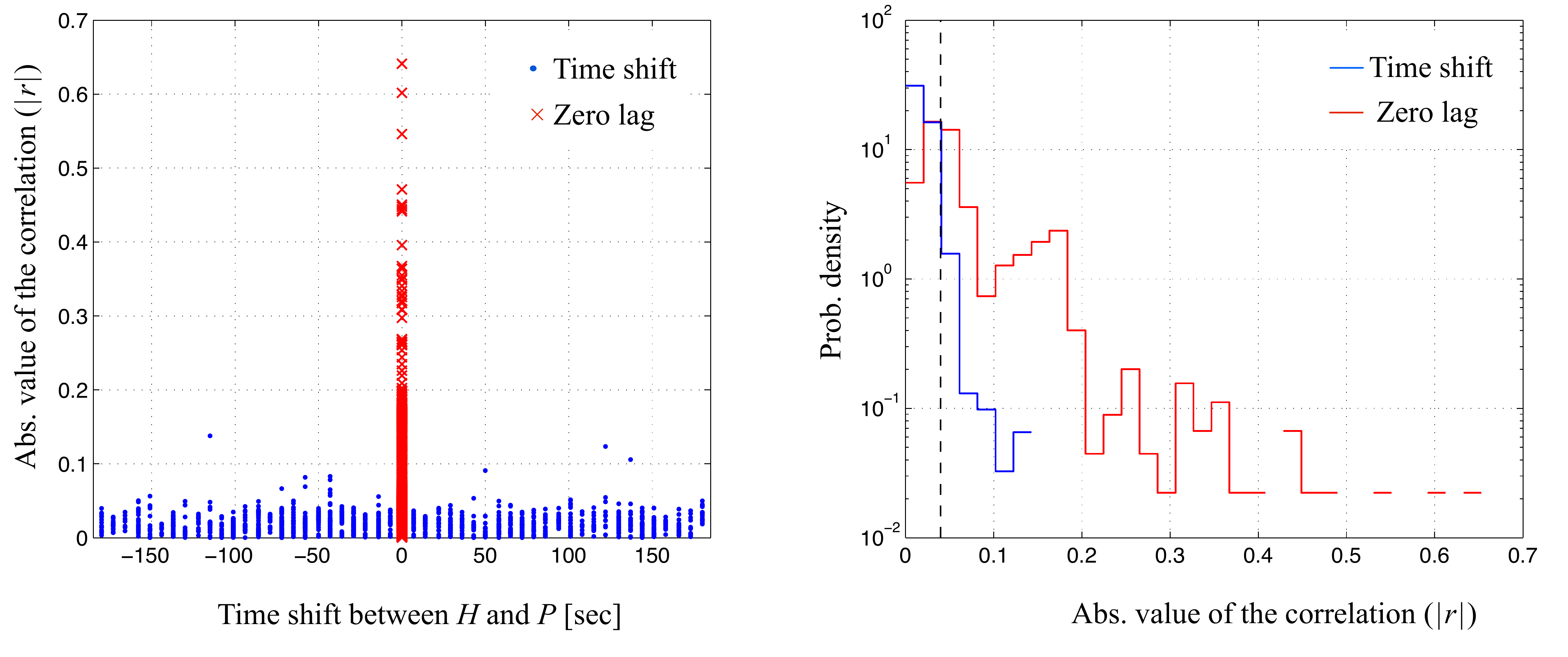}
\caption{An example of the correlation of a pseudo channel $P_{ij}$ with the GW channel $H$. 
The left panel shows the absolute value of the linear correlation coefficient $r$ between 
$H$ and $P_{ij}$ as a function of the time shift between 
the channels, while the right panel shows the distribution of $|r|$ from the time-shifted
coincidences (blue) and zero lag (red). The pseudo channel is constructed from a bilinear 
combination of {LSC-MICH\_CTRL} channel and {ASC-QPDY\_P} from the L1 detector.}
\label{fig:TimeShiftPlot}
\end{figure*}

``Traditional'' veto methods that look for time-coincidence between triggers in the GW channel and an 
auxiliary channel have been quite successful in reducing the rate of spurious triggers 
in the GW channel. However, more powerful veto methods can be formulated making use of the 
information contained in the time-series data recorded in the two channels as well as our 
empirical understanding of the coupling of different noise sources to the GW channel. 
An instrumental veto method making use of the knowledge of the coupling of different detector subsystems 
to the GW channel was proposed in Refs.~\cite{Ajith:2007hg,Ajith:2006ng}. The main idea behind 
this method was that the noise recorded by an instrumental channel \x\ can be \emph{transferred}~\cite{Adhikari:2006,smith2006linear} to 
the GW channel \h\ provided the noise coupling is \emph{linear} and the transfer function is 
known. This allows us to predict how a glitch witnessed by an auxiliary channel \x\ would appear in the 
GW channel \h. If a glitch is found in the GW channel at the same time and is consistent with 
the ``prediction'', then it can be vetoed with high confidence. This method was found 
to be a very efficient veto method in the fifth science run of GEO\,600~\cite{Ajith:2007hg}. 

Here we present a powerful veto method using a more complex, \emph{bilinear}-coupling model. 
The choice of the bilinear-coupling model is motivated by the empirical 
observations~\cite{Whitcomb:2000,Shoemaker:comm,Weiss:comm}
from the LIGO interferometers that glitches witnessed by auxiliary channels appear in the 
GW channel only at particular epochs of time, suggesting a time-varying transfer function
for the coupling of noise from the auxiliary channel to the GW channel. 
For example, glitches in the channel recording the signal controlling the length of
the power-recycling cavity are found to couple to the GW channel only during 
particular states of the detector. Signatures of such instabilities are recorded in 
many other channels.  For example, slow drifts in the alignment of mirrors can cause the position of the laser beam 
to wander around the mirror surfaces, which can potentially affect the coupling of the 
glitches in channel recording the Michelson control signal to the GW channel. Such slow, 
angular motions are recorded by the quadrant photodiodes (QPDs) placed behind the end-mirrors. This 
suggests that the signal recorded by the photodiode contains some information of the time-varying 
transfer function describing the coupling of the power-recycling control with the GW channel. 
A schematic diagram of the power-recycled Michelson interferometer is given in Figure~\ref{fig:LIGODiagram}.

In this paper, we denote the ``primary channels'' (channels recording fast motions such as glitches) 
by $X_i$ and ``secondary channels'' (those recording slow configuration changes such as  alignment 
drifts, slow angular motions of the beam, etc.) by $Y_j$.  Table~\ref{tab:bilinear_coupling_examples} 
lists some of the potential mechanisms in which a bilinear combination of ``primary'' and ``secondary'' 
channels couple to the GW channel. 

\begin{table*}
\centering
\begin{tabular}{l@{\quad}l}
\toprule
Primary channel $(X_i)$ & Secondary channel $(Y_i)$ \\
\midrule
Length feedback control signals & Beam position on the mirrors \\
(e.g., LSC-MICH\_CTRL, LSC-PRC\_CTRL) & (e.g., ASC-QPD\{X,Y\}\_\{P,Y\})\\
Angular torque feedback control signals & Beam position on the mirrors \\
(e.g., ASC-ETM\{X,Y\}\_\{P,Y\}) & (e.g., ASC-QPD\{X,Y\}\_\{P,Y\})\\
Length feedback control signals & Interferometer misalignment signals \\
(e.g., LSC-MICH\_CTRL, LSC-PRC\_CTRL) & (e.g., ASC-WFS\{1,2\}\_Q\{P,Y\})\\
\bottomrule
\end{tabular}
\caption{Examples of potential bilinear-coupling mechanisms in LIGO detectors. Bilinear combinations of the primary and 
secondary channels, $x_i(t) \, y_i(t)$, have been found to be correlated to the GW channel. See Section~\ref{sec:veto_analysis_s6}
for a description of the channel names.}
\label{tab:bilinear_coupling_examples}
\end{table*}

We consider a simple model of the coupling between a combination of instrumental 
channels and the GW channel. In particular, our hypothesis is that many of the glitches in the GW channel $H$ (with time series data $h(t)$)
can be ``best-witnessed'' by a bilinear combination of a primary channel $X_i$ and a secondary channel $Y_j$. \ie, 
\begin{equation}
h(t) \propto p_{ij}(t),
\end{equation}
where 
\begin{equation}
p_{ij}(t) \equiv  x_i(t)\,y_j(t) 
\label{eq:pseudo_channel_def}
\end{equation}
denote the data from a ``pseudo channel'' $P_{ij}$ which 
is a bilinear combination of $x_i(t)$ and $y_j(t)$, the time-series data recorded in $X_i$ and $Y_j$. 

The consistency of the glitches in the GW channel $H$ with the pseudo channel 
$P_{ij}$ can be tested by computing the linear correlation coefficient between $ p_{ij}(t)$ and $h(t)$, 
over an appropriate frequency band:
\be
r_{ij} \equiv \frac{\left< h \, , \, p_{ij} \right> }
{||h||~||p_{ij}||},
\ee
where the angular brackets denote inner products, and $||a|| \equiv \left<a,a\right>$ denotes the 
magnitude of the vector $a$. That is,
\begin{equation}
\left<a\,,b\right> \equiv \int_{f_\mathrm{min}}^{f_\mathrm{max}} \tilde{a}(f) \, \tilde{b}^*(f) \, df,
\end{equation}
where $\tilde{a}(f)$ is the Fourier transform of the time series data $a(t)$ over some time duration 
comparable to the duration of the glitch under consideration, and $f_\mathrm{min}$ and $f_\mathrm{max}$ 
are appropriately chosen low and high frequency cutoffs (e.g., 
from the bandwidth of the glitch under consideration, as estimated by a glitch detection algorithm)~
\footnote{In the work reported in Sec.~\ref{sec:veto_analysis_s6}, a fixed frequency range of 32--4096 Hz
was chosen as the burst detection algorithm KW does not estimate the bandwidth of the glitch.}. 

We compute the correlation coefficient $r_{ij}$ between $ p_{ij}(t)$ and $h(t)$ at the time of a 
coincident trigger in the primary instrumental channel $X_i$ and the GW channel $H$. If the trigger in 
$H$ is causally related to the one in $X_i$, and if our coupling model is realistic, we expect $|r_{ij}| \gg 0$.
On the other hand, if the coincidence of the triggers in $H$ and $X$ is expected to be purely accidental, we 
expect $|r_{ij}| \sim 0$. An appropriately determined threshold $\lambda_{ij}$ can be used to decide 
whether the correlation is significant. If $|r_{ij}| > \lambda_{ij}$, the trigger in $H$ can be  
vetoed. 

In order to determine an appropriate threshold, it is important to understand the ``background''
distribution of $r_{ij}$ -- the distribution of $r_{ij}$ arising from purely accidental correlations. 
In order to estimate the background distribution of $r_{ij}$, we time shift the data between 
the auxiliary channels and the GW channel (by an amount much larger than the correlation length 
of the data), so that all the real correlations between the auxiliary channels and the GW channel 
are removed. Any remaining correlations between the channel pairs are purely accidental. 
We then identify coincident noise transients between the channel pairs and compute $r_{ij}$ using
data surrounding the triggers. An example of such ``time-shift'' analysis is shown in 
Figure~\ref{fig:TimeShiftPlot}, along with the correlations in the ``zero lag'' (no time shift 
applied between the channel pairs so that the correlations are real). A suitable threshold on 
$r_{ij}$ can be found from the time-shifted analysis so that only an acceptable number of ``background
triggers'' have $r_{ij}$ greater than the threshold. This threshold can be used to decide which 
of the coincident triggers in the zero lag should be vetoed. 

\begin{figure}[tbh]
\centering
\includegraphics[width=2.5in]{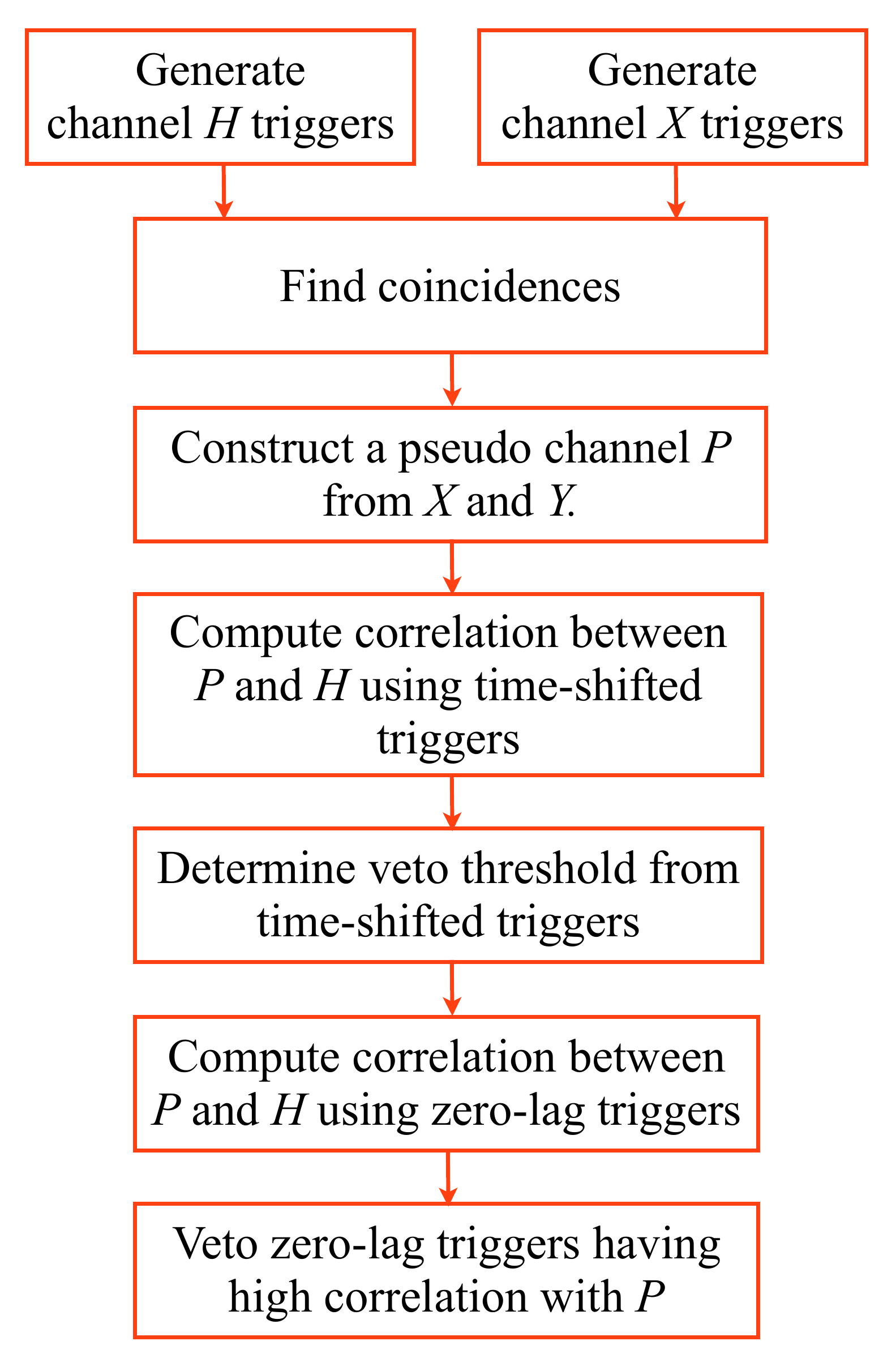}
\caption{Schematic diagram of the veto analysis pipeline.}
\label{fig:VetoPipeline}
\end{figure}

A schematic diagram of the vetoing  algorithm is given in Figure~\ref{fig:VetoPipeline}.
First we identify noise transients in the GW channel $H$ and one ``primary'' instrumental 
channel $X_i$ using an appropriate event trigger generator (such as KW). Coincident triggers between 
$H$ and $X_i$ are identified, allowing a time-window of the order of a second for coincidence.  
A pseudo channel $P_{ij}$ is 
constructed according to Eq.~(\ref{eq:pseudo_channel_def}) for a selected set of 
candidates for $Y_j$. The linear correlation coefficient $r_{ij}$ between $H$ and $P_{ij}$ is
computed at the time of each coincident trigger in zero lag and in each time shift. 
The distribution of $r_{ij}$ 
from the time-shifted analysis gives the background distribution of the correlation. A threshold 
$\lambda_{ij}$ is chosen such that only a very small fraction of the coincident triggers in the 
time shift have $r_{ij}$ greater than the threshold. After that, the analysis is repeated 
without applying any time shift between $H$ and $X_i$ (``zero-lag'' analysis), and all 
coincident triggers with $r_{ij} > \lambda_{ij}$ are vetoed. 

\section{Application to data from the sixth science run of LIGO}
\label{sec:veto_analysis_s6}

\begin{figure}[tbh]
\centering
\includegraphics[width=3.2in]{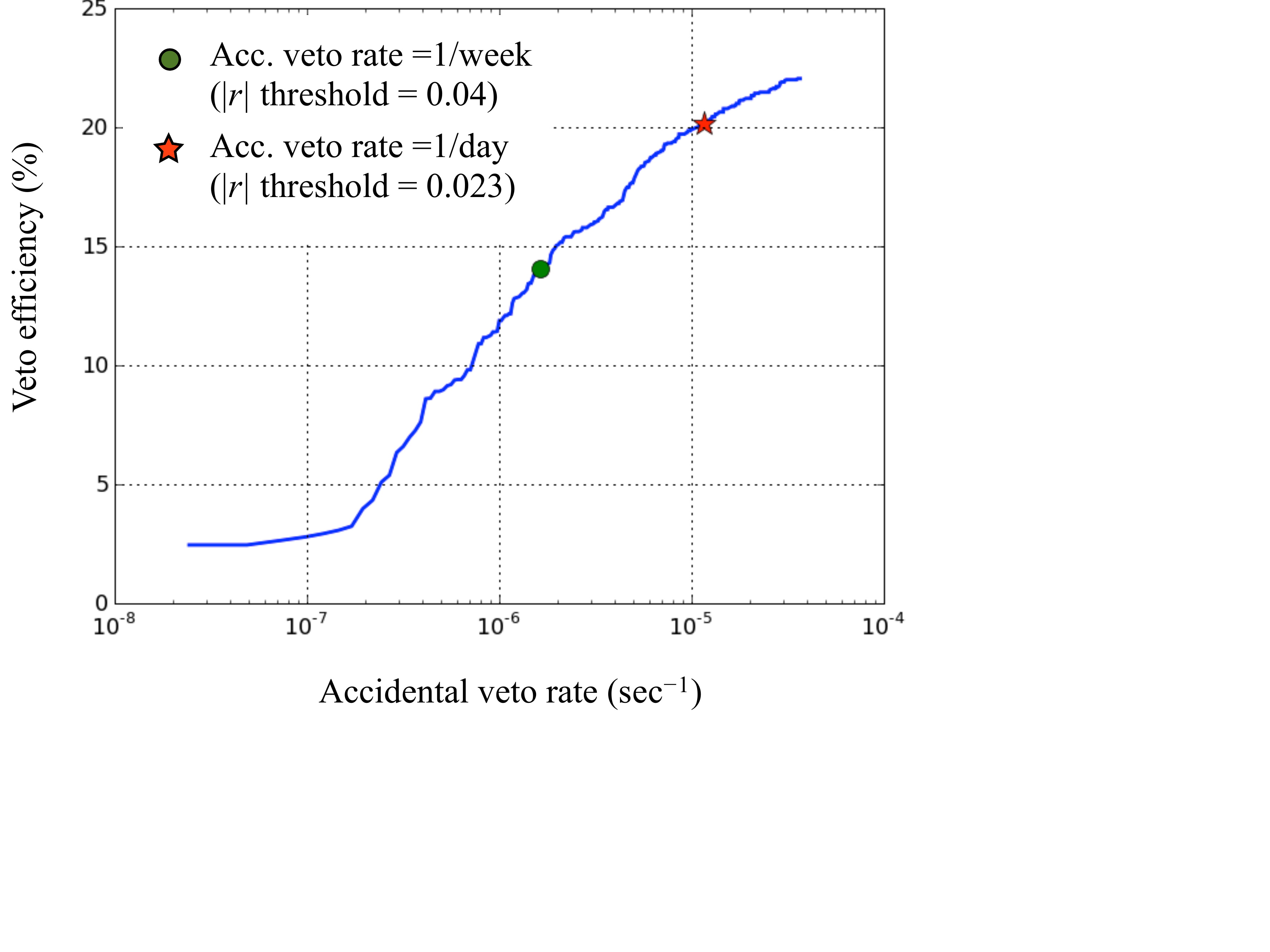}
\caption{The veto efficiency (fraction of triggers in $H$ vetoed in the zero lag) as a function 
of the accidental veto rate (fraction of triggers in $H$ vetoed per unit time in the time shifted analysis).
The pseudo channel for this example is constructed from a bilinear combination of {LSC-MICH\_CTRL} 
channel and {ASC-QPDY\_P} from the L1 detector. The veto efficiency and accidental 
veto rate are computed by changing the veto thresholds. The green circle and the red star 
correspond to accidental veto rates of 1 per week and 1 per day, respectively. Thresholds
corresponding to an accidental veto rate of 1 per week were used for the final veto analysis.}
\label{fig:ROCplot}
\end{figure}

\begin{table}
\centering
\begin{tabular}{l@{\quad}l}
\toprule
Primary channels $X_i$ & Secondary channels $Y_j$ \\  
\midrule 
ASC-ETMX\_\{P,Y\} & ASC-QPDX\_\{P,Y\} \\
ASC-ETMY\_\{P,Y\} & ASC-QPDY\_\{P,Y\} \\
ASC-ITMX\_\{P,Y\} & ASC-WFS1\_Q\{P,Y\} \\
ASC-ITMY\_\{P,Y\} & ASC-WFS2\_I\{P,Y\} \\
LSC-MICH\_CTRL    & ASC-WFS2\_Q\{P,Y\} \\
LSC-PRC\_CTRL     & ASC-WFS3\_I\{P,Y\} \\
				  & ASC-WFS4\_I\{P,Y\} \\
\bottomrule
\end{tabular}
\caption{Auxiliary channels used for the bilinear-coupling veto analysis in the S6 data. Pseudo channels were constructed using all the 140 bilinear combinations $x_i(t)\, y_j(t)$. See the text for a description of various channels.}
\label{tab:channel_list}
\end{table}

The sixth science run (S6) of the LIGO-Livingston (L1) and LIGO-Hanford (H1) detectors lasted from 7 July
2009 to 20 October 2010, 
and 141 days of coincident H1-L1 science quality data was collected during this run. Owing to a number
of improvements made on the LIGO detectors, the sensitivity of the detectors was in general better than
the previous science runs. However, the rate of non-Gaussian noise transients was found to be larger
than in the previous science runs. A number of instrumental veto techniques, as noted in Section~\ref{sec:intro}, 
were employed to reduce many of the noise transients in the data to be analyzed, and 
hence to improve the sensitivity of the searches for transient GW signals. Here we summarize 
the veto analysis performed on the S6 data using the bilinear-coupling veto pipeline.

The detector~\cite{Abbott:2007kv} auxiliary channels which we found to be most useful
for the bilinear veto analysis described in this paper  are listed in 
Table~\ref{tab:channel_list}. We chose a set of 10 primary (fast) and 14 secondary
(slow) auxiliary channels. In addition, we also perform the veto analysis using 
10 primary channels assuming a linear coupling model. 
These channels are all from the LIGO interferometer sensing 
and control (ISC) system. No physical environment monitoring (PEM) channels were 
used for this bilinear veto analysis.
The ISC channels are all derived from optical sensing of length
degrees of freedom (LSC) or angular degrees of freedom (ASC) of the
interferometer; the optical sensors are read out by (near-DC or radio frequency, RF)
photodiodes, analog electronics, and analog-to-digital conversion at
16384 Hz for the LSC signals and 2048 Hz for the ASC channels. A
digital control system is employed to maintain these degrees of
freedom at their nominal values throughout the observational run. The
most useful length sensing (LSC) channels, derived from readout of RF
photodiodes sensing the laser beam reflected from the interferometer~\cite{Fritschel:01}, 
were the one used to keep the power recycling cavity (PRC)
resonant (LSC-PRC\_CTRL), and the one used to keep the short Michelson (MICH)
length fixed (LSC-MICH\_CTRL). The most useful fast angular
sensing (ASC) channels, for both pitch (P) and yaw (Y) angular degrees
of freedom, were the ones monitoring the LIGO arm cavity mirrors,
which we refer to as test masses: the input test mass (ITM) and end test
mass (ETM) on each of the X and Y arms. So, for example, ASC-ETMY\_Y
represents the (optically-based) signal monitoring the yaw angular
degree of freedom of the end test mass on the interferometer's Y arm.
The most useful slow ASC channels were those from quadrant (near-DC)
photodiodes monitoring the light transmitted through the ETMs on both
arms (e.g., ASC-QPDX\_P is the signal monitoring the pitch angular degree
of freedom of the laser beam exiting the end of the X arm), and the
quadrant RF photodiodes (wavefront sensors, WFS) monitoring light
from the output ports of the interferometer to measure deviations from optimal
global alignment in six angular degrees of freedom~\cite{Fritschel:98}. So, for
example, ASC-WFS1\_QP represents the wavefront sensor signal monitoring
the pitch angular degree of freedom of the end test masses of the two
interferometer arms (differentially).

From the primary and secondary auxiliary channels listed in Table~\ref{tab:channel_list},
140 pseudo channels were constructed as bilinear combinations [see Eq.(\ref{eq:pseudo_channel_def})]. 
In addition, we also performed veto analysis using a linear coupling model using the 10 primary 
auxiliary channels (where, we threshold
on the correlation between the GW channel and the primary instrumental channels $X_i$). Transients 
in the primary auxiliary channels as well as the GW channel were detected using KW~\cite{KleineWelle}. 
We considered KW triggers with signal-to-noise ratio (SNR) greater
than 8. Coincidence triggers between the primary auxiliary and GW channels were identified using a time
window of 0.5 seconds, and the veto analysis was performed using each of the 140 pseudo channels (using
the bilinear-coupling model), and 10 primary channels (using the linear coupling model). 
Correlation coefficients for each set of coincident triggers were calculated as described in 
Section~\ref{sec:vetomethod}. The length of the data used to compute the correlation coefficient 
was chosen to be the cumulative duration (typically less than a second) of the coincident triggers 
in the two channels. Fifty time shifts in the interval [$-$180s, 180s] were performed for each channel pair to estimate the 
background distribution of the correlation coefficients. Thresholds estimated from this time shift analysis 
were used to veto coincident triggers in the zero lag. 
For choosing the thresholds on the correlation, we define some useful figures of merit. One is the \emph{efficiency}
of the veto, which is defined as the fraction of triggers in the GW channel vetoed in the zero lag. 
The second is the \emph{accidental veto rate}, which is the number of accidentally vetoed triggers in the 
GW channel per unit time (since they happen to be correlated with the pseudo channel purely by chance). We 
estimate the accidental veto rate by counting the fraction of triggers in $H$ vetoed per unit time in the 
time shifted analysis. Figure~\ref{fig:ROCplot} shows an example of the tuning used to determine the veto 
thresholds. The plot shows the \emph{veto efficiency} and the \emph{accidental veto rate} for different veto 
thresholds. As expected, higher thresholds result in lower veto efficiencies and lower accidental veto rates. 
In the final analysis we choose thresholds corresponding to accidental veto rate of 1 per week. Essentially 
this mean that, given the glitch rates in the GW channel and the auxiliary channel that we consider, our 
pipeline will veto a maximum of one trigger in the GW channel per week because it happens to have an 
accidental correlation (that is greater than the chosen threshold) with the particular pseudo channel 
under consideration. In addition to this, we also impose a threshold on the \emph{significance} of the veto, 
defined as the fraction of vetoed triggers in the zero lag divided by the fraction of vetoed triggers in the 
time-shift analysis. Only those pseudo channels for which significance is greater than 5 are used for 
vetoing triggers in the GW channel. Table~\ref{tab:params_summary} provides a summary of the parameters 
used in the analysis. 

Another important concern  is the \emph{safety probability} of the veto, which is the probability of 
vetoing an actual GW signal. In order to estimate the safety probability, we perform the veto analysis 
on the triggers generated from GW-like ``hardware injections'' (GW signals artificially injected to the 
hardware of the detector). Our estimate of the safety probability is 1$-$fraction of hardware injections 
vetoed. While this estimate is limited by the number and the nature of hardware injections performed, 
this gives us a reasonable estimate of the safety of the veto. ($\sim$ 3000 [2700] injections of compact 
binary coalescences and unmodelled burst signals were performed in H1 [L1] during S6, out of which $\sim$ 
2000 [1600] were detected by KW). For the S6 analysis, only those 
pseudo channels for which the safety probability is greater than 0.999 were used to veto triggers in the 
GW channel. However, we found that over the whole analysis from S6, none of the hardware injections were 
vetoed using any of the pseudo channels that we used. 

\begin{figure}[tb]
\centering
\includegraphics[width=3.5in]{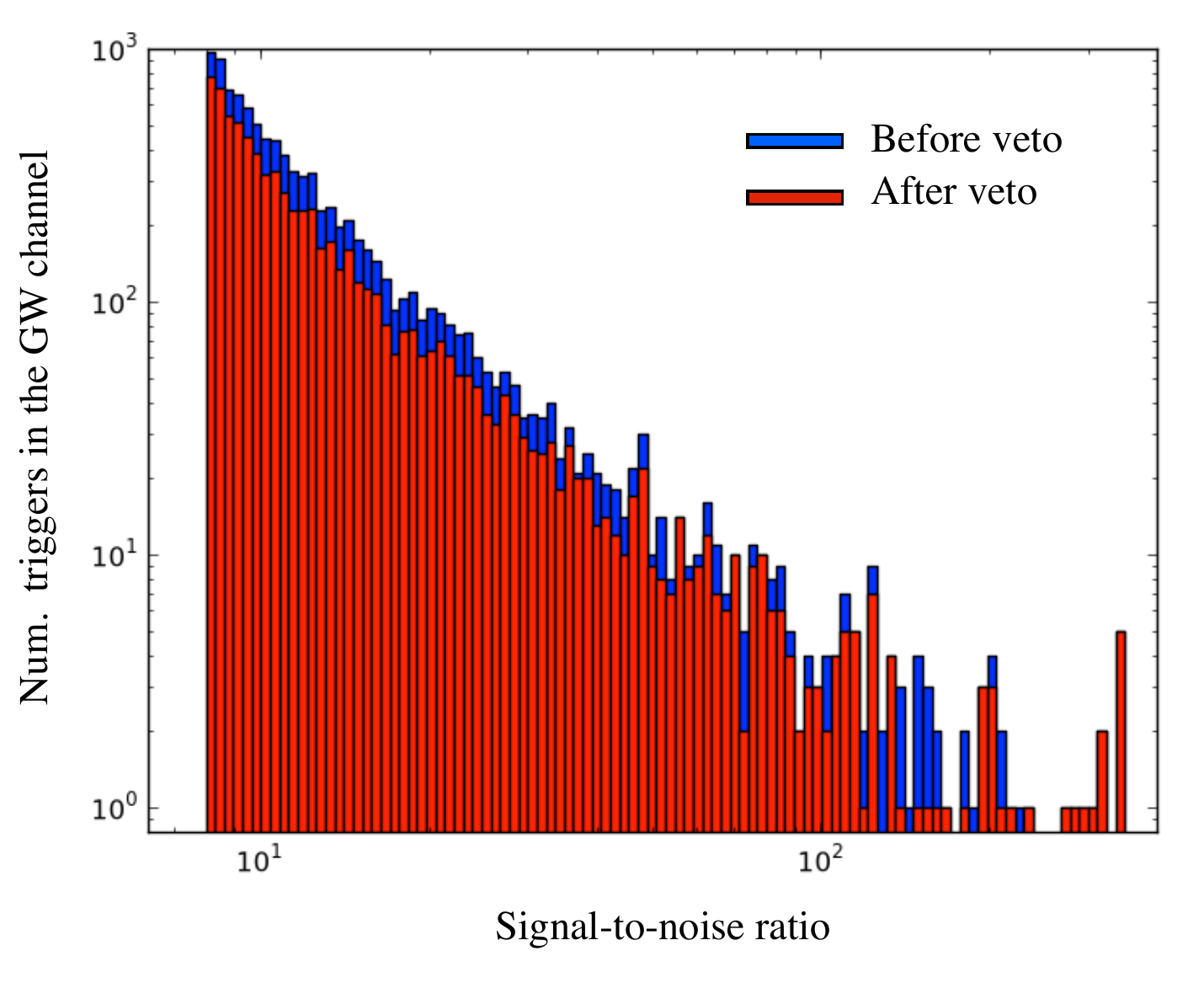}
\caption{The distribution of the signal-to-noise of the KW triggers in the GW channel before and after applying the bilinear-coupling veto. These triggers are from 1 week of L1 data starting from Oct 20 2009 01:12:42 UTC. Out of 9712 triggers in the GW channel, 2446 are vetoed.} 
\label{fig:veto_analysis_hist}
\end{figure}

\begin{figure}[tb]
\centering
\includegraphics[width=3.45in]{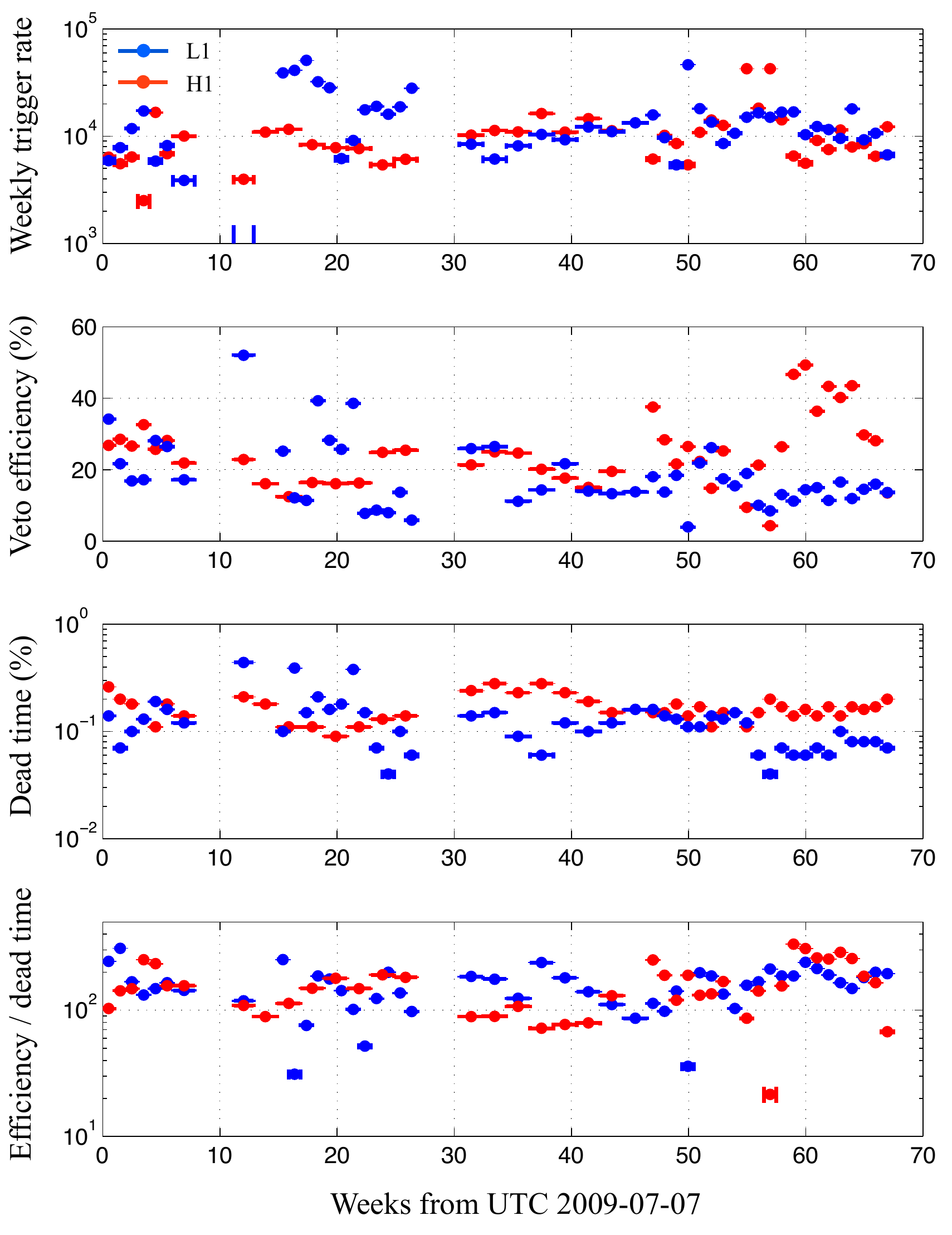}
\caption{A brief summary of the bilinear-coupling veto analysis performed on the S6 data in H1 and L1. The top panel shows the number of KW triggers from the GW channel per week, the second panel shows the veto efficiency (percentage of triggers vetoed), the third panel shows the dead time (percentage of observational data vetoed) and the bottom panel shows the ratio of the veto efficiency and dead time (a typical figure of merit used to quantify the effectiveness of a veto method).}
\label{fig:veto_analysis_summary}
\end{figure}

Figure~\ref{fig:veto_analysis_hist} shows the distribution of the SNR of the KW GW triggers 
before and after applying the veto, for a particular week during the S6 run in L1, while Fig.~\ref{fig:veto_analysis_summary} 
provides a quick summary of the bilinear-coupling veto analysis results generated in the S6 run.  
Figure~\ref{fig:veto_analysis_summary} shows the weekly glitch rate (defined as the number of KW triggers per 
week with SNR $>$ 8), the veto efficiency (fraction of triggers in the GW channel vetoed using all the 140 pseudo channels + 
10 linear coupling channels), the \emph{dead time} (fractional duration of the data that has been vetoed)
and the ratio of the veto efficiency and dead time (a typical figure of merit used to quantify the 
effectiveness of a veto method) over the entire S6 run in H1 and L1. In summary, the bilinear-coupling veto was 
found to be an efficient veto method with acceptable background rate, very low dead time and very high safety during the S6 
analysis. Along with other veto methods~\cite{hveto,UPV,MacLeod:2011up}, which also provided comparable veto efficiency, this veto was 
used to reduce the background rates of searches for transient GWs~\cite{S6inspiral,S6BHinspiral}.

\begin{table}
\centering
\begin{tabular}{l@{\quad}l}
\toprule
Parameter & Value \\
\midrule
Total number of (pseudo) channels & 140 bilinear + 10 linear \\
Accidental veto rate & 1 / week / pseudo channel \\
Threshold on veto significance & 5 \\
Threshold on trigger SNR & 8 \\
Threshold on safety probability & 0.999 \\
\bottomrule
\end{tabular}
\caption{Parameters for the bilinear-coupling veto analysis performed in the S6 data.}
\label{tab:params_summary}
\end{table}

\section{Future work: Bilinear veto as a potential interferometer diagnostic tool}
\label{sec:future_work}
The idea of the bilinear veto is to see whether a pseudo instrumental channel is correlated with the GW channel 
at glitchy times. The pseudo channel $P_{ij}$ is constructed as a bilinear combination of a ``primary" 
instrumental channel $X_i$ (which is producing glitches) and a ``secondary'' instrumental channel $Y_j$ which 
acts as a coupling agent, or a time-varying transfer function. Often, the secondary channels come as 
orthogonal pairs (such as the pitch and yaw of the movement of a mirror). Thus, the pseudo channels $P_{i1}$ 
and $P_{i2}$ constructed from two orthogonal secondary channels $Y_1$ and $Y_2$ contain fairly independent 
information. We can combine the correlation coefficients (a complex number) of the two pseudo channels 
into a single value. That is, we define
\begin{equation}
r_i \equiv \sqrt { |r_{i1}|^2 + |r_{i2}|^2 }, 
\end{equation}
where $r_1$, $r_2$ are the linear correlation coefficients of pseudo channels $P_{i1}$ and $P_{i2}$ with 
$H$. For certain channels, ($y_1(t), y_2(t)$) has a clear physical interpretation. For example, for the 
case of the secondary channels ASC-QPDX\_P and ASC-QPDY\_P this would correlate with the location 
of the beam spot on the two dimensional surface of the end-mirror of the Michelson cavity. If a significant 
fraction of the vetoed glitches are concentrated in a small region, that potentially suggest that the beam 
hitting on that particular position on the mirror makes it susceptible for the glitches in the the particular 
auxiliary channel to couple to the GW channel. This can be used as a potential diagnostic tool for the
commissioners to identify non-optimal detector states. 

\begin{figure}[tb]
\centering
\includegraphics[width=3.5in]{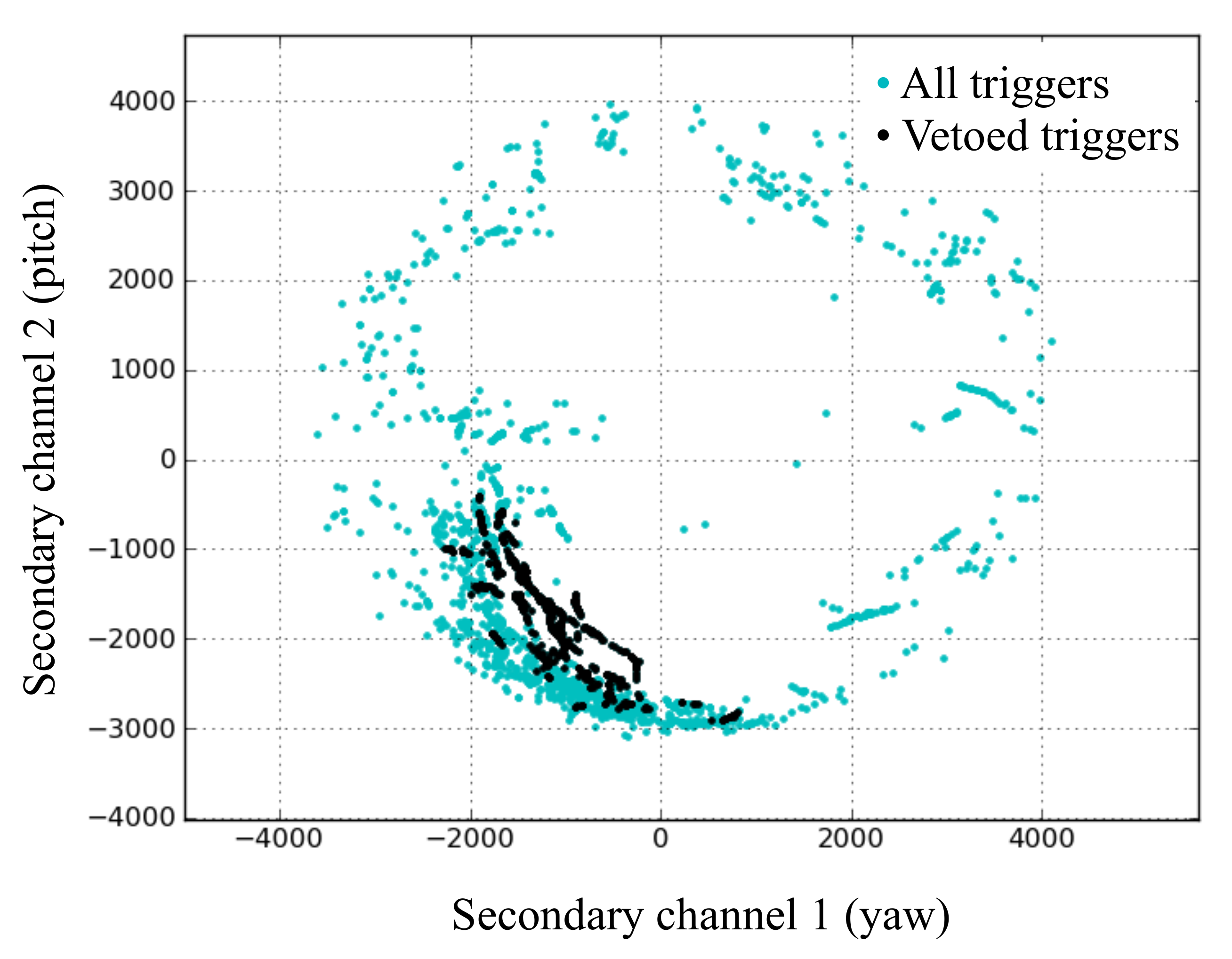}
\caption{The green dots correspond to the signal amplitude of the two secondary channels ASC-WFS2\_I\_{P} (pitch) 
and ASC-WFS2\_I\_{Y} (yaw) at the time of coincident glitches between L1-LSC-MICH\_CTRL and $H$, and the 
black dots correspond to the vetoed triggers among them. This plot suggests that glitches in 
L1-LSC-MICH\_CTRL couple to $H$ only during times when ASC-WFS2\_I\_{Pitch,Yaw} are in certain ``states''.}
\label{fig:detector_state}
\end{figure}

This aspect of the bilinear-coupling veto analysis was not investigated in detail during the S6 analysis. 
However, preliminary investigations suggest this as a promising diagnostic tool. 
An example is shown in Figure~\ref{fig:detector_state}, which plots the measured values of an orthogonal 
pair of secondary channels (ASC-WFS2\_I\_{P} and ASC-WFS2\_I\_{Y}) at the times of coincident triggers 
between the GW channel and the control signal to the Michelson cavity (from one day of data on 15 Nov 
2009). It can be seen that all the vetoed triggers (i.e., triggers in the GW channel that are highly 
correlated with the triggers in the pseudo channel under consideration) are clustered in a small region 
in the x-y plane. We plan to develop diagnostic tools based on more realistic bilinear-coupling models 
for the characterization of advanced GW detectors. 

\section{Conclusions}
\label{sec:conclusions}
In this paper we have presented a description of a novel veto method that was recently used to eliminate short duration noise transients (glitches) in data from the LIGO detectors during the S6 science run~\cite{S6inspiral,S6BHinspiral}. The unique aspect of the bilinear-coupling veto, as opposed to other vetoes used by LIGO and Virgo~\cite{hveto,UPV,PQveto}, is that it provides a means to identify and eliminate glitches in a detector's output GW channel that are associated with non-optimal states of interferometer sub-systems; these non-optimal states are observed in slow auxiliary channels, like the ones studied in this paper. This veto was also developed with the goal to see if the data from an interferometric detector's output GW strain channel at the time of an apparent signal is consistent with the data from a detector auxiliary channel, or a combination of auxiliary channels. Results were presented demonstrating the effectiveness of this veto with LIGO S6 data.

For the case of the upcoming advanced detectors like Advanced LIGO~\cite{aLIGO} and Advanced Virgo~\cite{aVirgo1,aVirgo2}, the severity of noise glitches in the GW strain channels is presently unknown. If such glitches are found to limit the ability to detect GW transient events, the the bilinear-coupling veto can be implemented as a means to reduce the number of noise transients. It should be noted, however, that Advanced LIGO and Advanced Virgo will reach their target sensitivities over a number of years of commissioning~\cite{Commissioning}. During this period it will be of critical importance to have tools that allow for the identification and characterization of noise. As demonstrated in this paper, the bilinear-coupling veto can be used as a means to diagnose sources of noise.

Another avenue for the improvement of the bilinear-coupling veto will be through the use of improved glitch trigger generators. The KW~\cite{KleineWelle} pipeline will continue to be used to generate triggers. However, new trigger pipelines with improved resolution at low frequencies are being developed. We also note that several other noise regression methods using linear/bilinear coupling models are being investigated within the LIGO-Virgo collaboration~\cite{Tiwari:2012,Klimenko:2012,Drago:2012}. We expect the bilinear-coupling veto to be a powerful noise diagnostic tool and veto generator for the next generation of laser interferometric GW detectors.

\acknowledgments 
We thank the LIGO Scientific Collaboration for allowing us to use the LIGO data used to conduct 
this study and Peter Shawhan for his comments on this manuscript. 
LIGO was constructed by the California Institute of Technology and Massachusetts Institute of 
Technology with funding from the National Science Foundation and operates under cooperative 
agreement PHY-0757058. PA's research was supported by the NSF grants PHY-0653653 and PHY-0601459, 
NSF career grant PHY-0956189, David and Barbara Groce Fund at Caltech, a FastTrack fellowship 
and a Ramanujan Fellowship from the Department of Science and Technology, India and by the EADS 
Foundation through a chair position on ``Mathematics of Complex Systems'' at ICTS-TIFR. NC's
research is supported by NSF grant PHY-1204371. This manuscript has the LIGO document number 
LIGO-P1400023-v2. 

\bibliography{BVeto}

\begin{thebibliography}{36}
\expandafter\ifx\csname natexlab\endcsname\relax\def\natexlab#1{#1}\fi
\expandafter\ifx\csname bibnamefont\endcsname\relax
  \def\bibnamefont#1{#1}\fi
\expandafter\ifx\csname bibfnamefont\endcsname\relax
  \def\bibfnamefont#1{#1}\fi
\expandafter\ifx\csname citenamefont\endcsname\relax
  \def\citenamefont#1{#1}\fi
\expandafter\ifx\csname url\endcsname\relax
  \def\url#1{\texttt{#1}}\fi
\expandafter\ifx\csname urlprefix\endcsname\relax\def\urlprefix{URL }\fi
\providecommand{\bibinfo}[2]{#2}
\providecommand{\eprint}[2][]{\url{#2}}

\bibitem[{\citenamefont{Abadie et~al.}(2012{\natexlab{a}})}]{S6inspiral}
\bibinfo{author}{\bibfnamefont{J.}~\bibnamefont{Abadie}} \bibnamefont{et~al.}
  (\bibinfo{collaboration}{LIGO Collaboration, Virgo Collaboration}),
  \bibinfo{journal}{Phys. Rev. D} \textbf{\bibinfo{volume}{85}},
  \bibinfo{pages}{082002} (\bibinfo{year}{2012}{\natexlab{a}}).

\bibitem[{\citenamefont{Aasi et~al.}(2013{\natexlab{a}})}]{S6BHinspiral}
\bibinfo{author}{\bibfnamefont{J.}~\bibnamefont{Aasi}} \bibnamefont{et~al.}
  (\bibinfo{collaboration}{LIGO Collaboration, Virgo Collaboration}),
  \bibinfo{journal}{Phys. Rev. D} \textbf{\bibinfo{volume}{87}},
  \bibinfo{pages}{022002} (\bibinfo{year}{2013}{\natexlab{a}}).

\bibitem[{\citenamefont{Abadie et~al.}(2012{\natexlab{b}})}]{S6burst}
\bibinfo{author}{\bibfnamefont{J.}~\bibnamefont{Abadie}} \bibnamefont{et~al.}
  (\bibinfo{collaboration}{LIGO Collaboration, Virgo Collaboration}),
  \bibinfo{journal}{Phys. Rev. D} \textbf{\bibinfo{volume}{85}},
  \bibinfo{pages}{122007} (\bibinfo{year}{2012}{\natexlab{b}}).

\bibitem[{\citenamefont{Ott et~al.}(2011)}]{Ott}
\bibinfo{author}{\bibfnamefont{C.}~\bibnamefont{Ott}} \bibnamefont{et~al.},
  \bibinfo{journal}{Physical Review Letters} \textbf{\bibinfo{volume}{106}},
  \bibinfo{pages}{161103} (\bibinfo{year}{2011}).

\bibitem[{\citenamefont{Aasi et~al.}(2013{\natexlab{b}})}]{S6string}
\bibinfo{author}{\bibfnamefont{J.}~\bibnamefont{Aasi}} \bibnamefont{et~al.}
  (\bibinfo{collaboration}{LIGO Collaboration, Virgo Collaboration})
  (\bibinfo{year}{2013}{\natexlab{b}}), \eprint{1310.2384}.

\bibitem[{\citenamefont{Aasi et~al.}(2012{\natexlab{a}})}]{S5CW}
\bibinfo{author}{\bibfnamefont{J.}~\bibnamefont{Aasi}} \bibnamefont{et~al.}
  (\bibinfo{collaboration}{LIGO Collaboration, Virgo Collaboration}),
  \bibinfo{journal}{Phys. Rev. D} \textbf{\bibinfo{volume}{87}},
  \bibinfo{pages}{042001} (\bibinfo{year}{2012}{\natexlab{a}}).

\bibitem[{\citenamefont{Abbott et~al.}(2009{\natexlab{a}})}]{S5stoch}
\bibinfo{author}{\bibfnamefont{B.}~\bibnamefont{Abbott}} \bibnamefont{et~al.}
  (\bibinfo{collaboration}{LIGO Collaboration, Virgo Collaboration}),
  \bibinfo{journal}{Nature} \textbf{\bibinfo{volume}{460}},
  \bibinfo{pages}{990} (\bibinfo{year}{2009}{\natexlab{a}}).

\bibitem[{\citenamefont{Abbott et~al.}(2012)}]{S6gamma}
\bibinfo{author}{\bibfnamefont{B.}~\bibnamefont{Abbott}} \bibnamefont{et~al.},
  \bibinfo{journal}{\apj} \textbf{\bibinfo{volume}{760}}, \bibinfo{pages}{12}
  (\bibinfo{year}{2012}).

\bibitem[{\citenamefont{Adrian-Martinez et~al.}(2013)}]{S5neutrinos}
\bibinfo{author}{\bibfnamefont{S.}~\bibnamefont{Adrian-Martinez}}
  \bibnamefont{et~al.}, \bibinfo{journal}{Journal of Cosmology and
  Astroparticle Physics} \textbf{\bibinfo{volume}{6}}, \bibinfo{pages}{008}
  (\bibinfo{year}{2013}).

\bibitem[{\citenamefont{Harry}(2010)}]{aLIGO}
\bibinfo{author}{\bibfnamefont{G.~M.} \bibnamefont{Harry}}
  (\bibinfo{collaboration}{LIGO Collaboration, Virgo Collaboration}),
  \bibinfo{journal}{Classical and Quantum Gravity}
  \textbf{\bibinfo{volume}{27}}, \bibinfo{pages}{084006}
  (\bibinfo{year}{2010}).

\bibitem[{\citenamefont{Acernese et~al.}(2009)}]{aVirgo1}
\bibinfo{author}{\bibfnamefont{F.}~\bibnamefont{Acernese}} \bibnamefont{et~al.}
  (\bibinfo{collaboration}{Virgo Collaboration}) (\bibinfo{year}{2009}),
  \urlprefix\url{https://tds.ego-gw.it/ql/?c=6589}.

\bibitem[{\citenamefont{Acernese et~al.}(2012)}]{aVirgo2}
\bibinfo{author}{\bibfnamefont{F.}~\bibnamefont{Acernese}} \bibnamefont{et~al.}
  (\bibinfo{collaboration}{Virgo Collaboration}) (\bibinfo{year}{2012}),
  \urlprefix\url{https://tds.ego-gw.it/ql/?c=8940}.

\bibitem[{\citenamefont{Aasi et~al.}(2013{\natexlab{c}})}]{Commissioning}
\bibinfo{author}{\bibfnamefont{J.}~\bibnamefont{Aasi}} \bibnamefont{et~al.}
  (\bibinfo{collaboration}{LIGO Collaboration, Virgo Collaboration})
  (\bibinfo{year}{2013}{\natexlab{c}}), \eprint{1304.0670}.

\bibitem[{\citenamefont{Somiya et~al.}(2012)}]{KAGRA}
\bibinfo{author}{\bibfnamefont{K.}~\bibnamefont{Somiya}} \bibnamefont{et~al.},
  \bibinfo{journal}{Classical Quantum Gravity} \textbf{\bibinfo{volume}{29}},
  \bibinfo{pages}{124007} (\bibinfo{year}{2012}).

\bibitem[{\citenamefont{Christensen}(2010)}]{S6DQ}
\bibinfo{author}{\bibfnamefont{N.}~\bibnamefont{Christensen}}
  (\bibinfo{collaboration}{LIGO Collaboration, Virgo Collaboration}),
  \bibinfo{journal}{Classical and Quantum Gravity}
  \textbf{\bibinfo{volume}{27}}, \bibinfo{pages}{194010}
  (\bibinfo{year}{2010}).

\bibitem[{\citenamefont{Aasi et~al.}(2012{\natexlab{b}})}]{VSR23DQ}
\bibinfo{author}{\bibfnamefont{J.}~\bibnamefont{Aasi}} \bibnamefont{et~al.}
  (\bibinfo{collaboration}{LIGO Collaboration, Virgo Collaboration}),
  \bibinfo{journal}{Classical and Quantum Gravity}
  \textbf{\bibinfo{volume}{29}}, \bibinfo{pages}{155002}
  (\bibinfo{year}{2012}{\natexlab{b}}).

\bibitem[{\citenamefont{Coughlin}(2010)}]{noise-lines}
\bibinfo{author}{\bibfnamefont{M.}~\bibnamefont{Coughlin}}
  (\bibinfo{collaboration}{LIGO Collaboration, Virgo Collaboration}),
  \bibinfo{journal}{J.Phys.Conf.Ser.} \textbf{\bibinfo{volume}{243}},
  \bibinfo{pages}{012010} (\bibinfo{year}{2010}).

\bibitem[{\citenamefont{Chatterji et~al.}(2004)\citenamefont{Chatterji,
  Blackburn, Martin, and Katsavounidis}}]{KleineWelle}
\bibinfo{author}{\bibfnamefont{S.}~\bibnamefont{Chatterji}},
  \bibinfo{author}{\bibfnamefont{L.}~\bibnamefont{Blackburn}},
  \bibinfo{author}{\bibfnamefont{G.}~\bibnamefont{Martin}}, \bibnamefont{and}
  \bibinfo{author}{\bibfnamefont{E.}~\bibnamefont{Katsavounidis}},
  \bibinfo{journal}{Classical and Quantum Gravity}
  \textbf{\bibinfo{volume}{21}}, \bibinfo{pages}{S1809} (\bibinfo{year}{2004}).

\bibitem[{\citenamefont{Smith et~al.}(2011)}]{hveto}
\bibinfo{author}{\bibfnamefont{J.}~\bibnamefont{Smith}} \bibnamefont{et~al.},
  \bibinfo{journal}{Classical and Quantum Gravity}
  \textbf{\bibinfo{volume}{28}}, \bibinfo{pages}{235005}
  (\bibinfo{year}{2011}).

\bibitem[{\citenamefont{Isogai}(2010)}]{UPV}
\bibinfo{author}{\bibfnamefont{T.}~\bibnamefont{Isogai}}
  (\bibinfo{collaboration}{LIGO Collaboration, Virgo Collaboration}),
  \bibinfo{journal}{J.Phys.Conf.Ser.} \textbf{\bibinfo{volume}{243}},
  \bibinfo{pages}{012005} (\bibinfo{year}{2010}).

\bibitem[{\citenamefont{MacLeod et~al.}(2012)\citenamefont{MacLeod, Fairhurst,
  Hughey, Lundgren, Pekowsky et~al.}}]{MacLeod:2011up}
\bibinfo{author}{\bibfnamefont{D.}~\bibnamefont{MacLeod}},
  \bibinfo{author}{\bibfnamefont{S.}~\bibnamefont{Fairhurst}},
  \bibinfo{author}{\bibfnamefont{B.}~\bibnamefont{Hughey}},
  \bibinfo{author}{\bibfnamefont{A.}~\bibnamefont{Lundgren}},
  \bibinfo{author}{\bibfnamefont{L.}~\bibnamefont{Pekowsky}},
  \bibnamefont{et~al.}, \bibinfo{journal}{Class.Quant.Grav.}
  \textbf{\bibinfo{volume}{29}}, \bibinfo{pages}{055006}
  (\bibinfo{year}{2012}), \eprint{1108.0312}.

\bibitem[{\citenamefont{Ballinger}(2009)}]{PQveto}
\bibinfo{author}{\bibfnamefont{T.}~\bibnamefont{Ballinger}}
  (\bibinfo{collaboration}{LIGO Collaboration, Virgo Collaboration}),
  \bibinfo{journal}{Classical and Quantum Gravity}
  \textbf{\bibinfo{volume}{26}}, \bibinfo{pages}{204003}
  (\bibinfo{year}{2009}).

\bibitem[{\citenamefont{Hewitson and Ajith}(2005)}]{Hewitson:2005wr}
\bibinfo{author}{\bibfnamefont{M.}~\bibnamefont{Hewitson}} \bibnamefont{and}
  \bibinfo{author}{\bibfnamefont{P.}~\bibnamefont{Ajith}},
  \bibinfo{journal}{Class.Quant.Grav.} \textbf{\bibinfo{volume}{22}},
  \bibinfo{pages}{4903} (\bibinfo{year}{2005}).

\bibitem[{\citenamefont{Ajith et~al.}(2007)}]{Ajith:2007hg}
\bibinfo{author}{\bibfnamefont{P.}~\bibnamefont{Ajith}} \bibnamefont{et~al.},
  \bibinfo{journal}{Phys. Rev. D} \textbf{\bibinfo{volume}{76}},
  \bibinfo{pages}{042004} (\bibinfo{year}{2007}), \eprint{0705.1111}.

\bibitem[{\citenamefont{Ajith et~al.}(2006)\citenamefont{Ajith, Hewitson,
  Smith, and Strain}}]{Ajith:2006ng}
\bibinfo{author}{\bibfnamefont{P.}~\bibnamefont{Ajith}},
  \bibinfo{author}{\bibfnamefont{M.}~\bibnamefont{Hewitson}},
  \bibinfo{author}{\bibfnamefont{J.}~\bibnamefont{Smith}}, \bibnamefont{and}
  \bibinfo{author}{\bibfnamefont{K.}~\bibnamefont{Strain}},
  \bibinfo{journal}{Class.Quant.Grav.} \textbf{\bibinfo{volume}{23}},
  \bibinfo{pages}{5825} (\bibinfo{year}{2006}), \eprint{gr-qc/0605079}.

\bibitem[{\citenamefont{Adhikari}(2004)}]{Adhikari:2006}
\bibinfo{author}{\bibfnamefont{R.}~\bibnamefont{Adhikari}}, Ph.D. thesis,
  \bibinfo{school}{Massachusetts Institute of Technology}
  (\bibinfo{year}{2004}).

\bibitem[{\citenamefont{Smith et~al.}(2006)\citenamefont{Smith, Ajith, Grote,
  Hewitson, Hild, L{\"u}ck, Strain, Willke, Hough, and
  Danzmann}}]{smith2006linear}
\bibinfo{author}{\bibfnamefont{J.}~\bibnamefont{Smith}},
  \bibinfo{author}{\bibfnamefont{P.}~\bibnamefont{Ajith}},
  \bibinfo{author}{\bibfnamefont{H.}~\bibnamefont{Grote}},
  \bibinfo{author}{\bibfnamefont{M.}~\bibnamefont{Hewitson}},
  \bibinfo{author}{\bibfnamefont{S.}~\bibnamefont{Hild}},
  \bibinfo{author}{\bibfnamefont{H.}~\bibnamefont{L{\"u}ck}},
  \bibinfo{author}{\bibfnamefont{K.}~\bibnamefont{Strain}},
  \bibinfo{author}{\bibfnamefont{B.}~\bibnamefont{Willke}},
  \bibinfo{author}{\bibfnamefont{J.}~\bibnamefont{Hough}}, \bibnamefont{and}
  \bibinfo{author}{\bibfnamefont{K.}~\bibnamefont{Danzmann}},
  \bibinfo{journal}{Classical and Quantum Gravity}
  \textbf{\bibinfo{volume}{23}}, \bibinfo{pages}{527} (\bibinfo{year}{2006}).

\bibitem[{\citenamefont{Whitcomb}(2000)}]{Whitcomb:2000}
\bibinfo{author}{\bibfnamefont{S.}~\bibnamefont{Whitcomb}},
  \emph{\bibinfo{title}{``{B}i-linear'' noise mechanisms in interferometers}}
  (\bibinfo{year}{2000}), \bibinfo{note}{presentation at GWDAW},
  \eprint{LIGO-G000336-00-D},
  \urlprefix\url{https://dcc.ligo.org/LIGO-G000336/public}.

\bibitem[{\citenamefont{Shoemaker}()}]{Shoemaker:comm}
\bibinfo{author}{\bibfnamefont{D.}~\bibnamefont{Shoemaker}},
  \bibinfo{note}{private communication}.

\bibitem[{\citenamefont{Weiss}()}]{Weiss:comm}
\bibinfo{author}{\bibfnamefont{R.}~\bibnamefont{Weiss}}, \bibinfo{note}{private
  communication}.

\bibitem[{\citenamefont{Abbott et~al.}(2009{\natexlab{b}})}]{Abbott:2007kv}
\bibinfo{author}{\bibfnamefont{B.}~\bibnamefont{Abbott}} \bibnamefont{et~al.}
  (\bibinfo{collaboration}{LIGO Scientific Collaboration}),
  \bibinfo{journal}{Rept.Prog.Phys.} \textbf{\bibinfo{volume}{72}},
  \bibinfo{pages}{076901} (\bibinfo{year}{2009}{\natexlab{b}}),
  \eprint{0711.3041}.

\bibitem[{\citenamefont{Fritschel et~al.}(2001)\citenamefont{Fritschel, Bork,
  Gonz\'{a}lez, Mavalvala, Ouimette, Rong, Sigg, and Zucker}}]{Fritschel:01}
\bibinfo{author}{\bibfnamefont{P.}~\bibnamefont{Fritschel}},
  \bibinfo{author}{\bibfnamefont{R.}~\bibnamefont{Bork}},
  \bibinfo{author}{\bibfnamefont{G.}~\bibnamefont{Gonz\'{a}lez}},
  \bibinfo{author}{\bibfnamefont{N.}~\bibnamefont{Mavalvala}},
  \bibinfo{author}{\bibfnamefont{D.}~\bibnamefont{Ouimette}},
  \bibinfo{author}{\bibfnamefont{H.}~\bibnamefont{Rong}},
  \bibinfo{author}{\bibfnamefont{D.}~\bibnamefont{Sigg}}, \bibnamefont{and}
  \bibinfo{author}{\bibfnamefont{M.}~\bibnamefont{Zucker}},
  \bibinfo{journal}{Appl. Opt.} \textbf{\bibinfo{volume}{40}},
  \bibinfo{pages}{4988} (\bibinfo{year}{2001}).

\bibitem[{\citenamefont{Fritschel et~al.}(1998)\citenamefont{Fritschel,
  Mavalvala, Shoemaker, Sigg, Zucker, and Gonz\'{a}lez}}]{Fritschel:98}
\bibinfo{author}{\bibfnamefont{P.}~\bibnamefont{Fritschel}},
  \bibinfo{author}{\bibfnamefont{N.}~\bibnamefont{Mavalvala}},
  \bibinfo{author}{\bibfnamefont{D.}~\bibnamefont{Shoemaker}},
  \bibinfo{author}{\bibfnamefont{D.}~\bibnamefont{Sigg}},
  \bibinfo{author}{\bibfnamefont{M.}~\bibnamefont{Zucker}}, \bibnamefont{and}
  \bibinfo{author}{\bibfnamefont{G.}~\bibnamefont{Gonz\'{a}lez}},
  \bibinfo{journal}{Appl. Opt.} \textbf{\bibinfo{volume}{37}},
  \bibinfo{pages}{6734} (\bibinfo{year}{1998}).

\bibitem[{\citenamefont{Tiwari et~al.}(2012)}]{Tiwari:2012}
\bibinfo{author}{\bibfnamefont{V.}~\bibnamefont{Tiwari}} \bibnamefont{et~al.},
  \emph{\bibinfo{title}{Regression of linear and bi-linear noise in {LIGO}}}
  (\bibinfo{year}{2012}), \bibinfo{note}{{LIGO} Internal Document},
  \eprint{LIGO-G1200288-v1}.

\bibitem[{\citenamefont{Klimenko}(2012)}]{Klimenko:2012}
\bibinfo{author}{\bibfnamefont{S.}~\bibnamefont{Klimenko}},
  \emph{\bibinfo{title}{Regression of {LIGO}/{Virgo} data}}
  (\bibinfo{year}{2012}), \bibinfo{note}{{LIGO} Internal Document},
  \eprint{LIGO-G1200197-v1}.

\bibitem[{\citenamefont{Drago}(2012)}]{Drago:2012}
\bibinfo{author}{\bibfnamefont{M.}~\bibnamefont{Drago}},
  \emph{\bibinfo{title}{Regression of {LIGO/Virgo} data}}
  (\bibinfo{year}{2012}), \bibinfo{note}{{LIGO} Internal Document},
  \eprint{LIGO-G1200278-v6}.

\end{thebibliography}

\end{document}